\documentclass{aa}  

\usepackage{color}
\usepackage{graphicx}
\usepackage{txfonts}
\begin{document}

   \title{Cosmic-ray production from neutron escape in microquasar jets}

   \subtitle{}

   \author{G.J. Escobar
          \inst{1}
          \and
          L.J. Pellizza\inst{2}
          \and
          G.E. Romero\inst{1}} 

   \institute{Instituto Argentino de Radioastronom\'ia, CONICET-CICBA, Argentina \and Instituto de Astronom\'ia y F\'isica del Espacio (IAFE), Universidad de Buenos Aires (UBA), Argentina\\
              \email{gescobar@iar.unlp.edu.ar}}

   \date{}

 
  \abstract
   {The origin of Galactic cosmic rays remains a matter of debate, but supernova remnants are commonly considered to be the main place where  high-energy cosmic rays are accelerated. Nevertheless, current models predict cosmic-ray spectra that do not match observations and the efficiency of the acceleration mechanism is still undetermined. On the other hand, the contribution of other kinds of sources to the Galactic cosmic-ray population is still unclear, and merits investigation.}
   {In this work we explore a novel mechanism through which microquasars might produce cosmic rays. In this scenario, microquasar jets generate  relativistic neutrons, which escape and decay outside the system; protons and electrons, created when these neutrons decay, escape to the interstellar medium as cosmic rays.}
   {We introduce the relativistic neutron component through a coupling term in the transport equation that governs the jet proton population. We compute the escape rate and decay distribution of these neutrons, and follow the propagation of the decay products until they escape the system and become cosmic rays. We then compute the spectra of these cosmic rays.}
   {Neutrons can drain only a small fraction of the jet power as cosmic rays. The most promising scenarios arise in extremely luminous systems ($L_\mathrm{jet} \sim 10^{40}\,\mathrm{erg \, s}^{-1}$), in which the fraction of jet power deposited in cosmic rays can reach $\sim 0.001$. Slow jets ($\Gamma \lesssim 2$, where $\Gamma$ is the bulk Lorentz factor) favour neutron production. The resulting cosmic-ray spectrum is similar for protons and electrons, which share the power in the ratio given by neutron decay. The spectrum peaks at roughly half the minimum energy of the relativistic protons in the jet; it is soft (spectral index $\sim 3$) above this energy, and almost flat below.}
   {The proposed mechanism produces more energetic cosmic rays from microquasars than those presented by previous works in which the particles escape through the jet terminal shock. Values of spectral index steeper than $2$ are possible for cosmic rays in our model and these indeed agree with those required to explain the spectral signatures of Galactic cosmic rays, although only the most extreme microquasars provide power comparable to that of a typical supernova remnant. The mechanism explored in this work may provide stronger and softer cosmic-ray sources in the early Universe, and therefore contribute to the heating and reionisation of the intergalactic medium.}

   \keywords{cosmic rays --- ISM: jets and outflows --- relativistic processes}

   \maketitle
%

\section{Introduction}

   Microquasars (MQs) are X-ray binaries (XRBs) with relativistic jets. These systems display a phenomenology that resembles that of active galactic nuclei (AGNs) but on smaller scales  \citep{Mirabel1994}. As jets inject large amounts of energy into the interstellar region, they are expected to influence the surrounding medium. \citet{Tetarenko2018, Tetarenko2020} present the detection of several transition spectral lines, which supports the presence of an interaction region between MQ jets and the interstellar medium (ISM). However, the authors find that the interaction region lies at much smaller scales than postulated in previous works.
    Microquasar jets may also heat the ISM by injecting kinetic energy. This issue has been investigated by \citet{Fender2005}, who conclude that the input of kinetic energy into the ISM from MQ jets has a minor, but non-negligible contribution with respect to supernovae.
    
    In addition, MQ jets could develop shocks that accelerate particles, injecting cosmic rays (CRs) into the Galaxy. Currently, the most plausible scenario is that supernova remnants (SNRs) accelerate CRs up to high energies. However, supernova shocks are in general non-relativistic, in contrast with the relativistic shocks in MQ jets. Thus, the spectrum of CRs produced in jets might be harder (i.e. with a larger fraction of high-energy CRs) than that of SNRs, and merits investigation. 
    
    \citet{Fender2005} estimate that $5-10$ per cent of the CR power of the Galaxy might be produced in the MQ population. Likewise, \citet{Heinz&Sunyaev2002} model a mechanism in which a narrow band of the CR spectrum is produced in the terminal shocks of MQ jets. This sharp spectrum has a characteristic proton energy that depends on the bulk Lorentz factor of the jet. \citet{Heinz&Sunyaev2002} also argue that a broader CR spectrum may arise from the collective effect of MQs with different Lorentz factors. 
    In a recent study, \citet{Cooper2020} reinforce the hypothesis of MQ jets as potential sources of CRs, suggesting that, because the maximum energy of jet-accelerated CRs is relatively high compared to other CR sources, the former may contribute significantly in the region between the Galactic and extragalactic components of the CR spectrum, that is, beyond the knee and below the ankle of this spectrum.
    
   To inject CRs in the Galaxy, the scenarios devised in the aforementioned works \citep{Heinz&Sunyaev2002, Fender2005, Cooper2020} require the presence of hadronic matter in MQ jets. This is supported by some theoretical models \citep[e.g.][]{Blandford&Payne1982}, in which a large-scale magnetic field launches an outflow of material from the accretion disc by a magneto-hydrodynamic mechanism. This leads to
   jets composed of a hot plasma of thermal electrons and hadrons, with a relativistic component. On the other hand, some MQs were found to have hadronic content.
   \citet{Migliari2002} made observations of the jet of SS~433, revealing the presence of 
   iron emission lines. Another case is the binary system 4U1630-47, for which 
   \citet{DiazTrigo2013} reported the detection of emission lines arising from baryonic matter travelling in the jet. Entrainment of matter from the stellar wind can also contribute to loading the jet with baryons \citep{Romero2003}.
   
   In this paper, we explore a different mechanism by which MQ jets may inject kinetic energy and CRs into the ISM. Unlike previous models, in which CRs are produced at the terminal shock of the jet, we propose that CRs may be injected directly into the ISM by relativistic neutrons escaping from the jet. These neutrons are produced by the interaction of protons accelerated in internal shocks with thermal ones. Neutrons escape freely from the system because they do not interact with the magnetic field confining the plasma. They decay far away from the jet, injecting relativistic protons and electrons into the ISM. 
   
   The population of CRs produced in our scenario may differ from that of previous models for several reasons. Firstly, escaping neutrons carry almost all the energy of their relativistic proton progenitors, whilst particles in the jet undergo adiabatic and radiation losses in their way to the termination shock. For the same reason, the CR energy distribution of our model may be different from that of particles escaping through the termination shock. Secondly, the injection of CRs proceeds at distances much greater than jet scales. Thirdly, CR injection by neutrons is roughly isotropic, because MQ jets have small Lorentz factors. Finally, electrons arising in neutron decay may be more energetic than those accelerated at the termination shocks. This is because the energy distribution of the neutron population is related to that of the relativistic proton population in the jet, which is subject to smaller losses and therefore attains higher maximum energies. 

  Relativistic neutron production has already been explored  in the jets of AGNs \citep{Kirk1989, Sikora1989, Giovanoni1990, Atoyan1992a, Atoyan1992b, Atoyan2003}. 
  The latter authors compute the photo-meson production of relativistic neutrons and the $\gamma$-ray production from interactions of these neutrons with photon fields. They show that the production of narrow beams of ultra high-energy neutrons with $E \gtrsim 10^{17}\,\mathrm{eV}$ is possible. The existence of this neutron component has not  yet been investigated in MQ jets. Such a population of relativistic neutrons may not only be a source of CRs, but may also have effects on the $\gamma$-ray emission of the jet.
   
This paper is organised as follows. In Sect.~\ref{Model} we introduce the jet model, describing the relevant physical processes that drive neutron production. In Sect.~\ref{sec:neutron_production} we use a fiducial set of parameters to apply the jet model. In addition, we compute the $\gamma$-ray spectrum of a typical MQ jet populated with protons and neutrons, and assess the observability of the neutron component. In Sect.~\ref{sec:cr_production} we follow neutron escape until decay, determining the power injected in CRs and their energy distribution. Finally, in Sect.~\ref{Conclusions} we discuss our results and present our conclusions.
   

\section{Jet model}
\label{Model}

   Our jet model is based on that of \citet{Romero&Vila2008}. Throughout this paper we use a fiducial set of parameters (summarised in Table~\ref{tab:modelparameters}) taken from Model A of \citet{Pepe2015}. \citet{Pepe2015} model the electromagnetic output of the microquasar \object{Cygnus X-1} accounting for the contribution of the companion star, jet, accretion disc, and a hot corona, and obtain two sets of parameters (models A and B) for their jet model that represent the best fit to the spectral energy distribution (SED) of \object{Cygnus X-1}. The uncertainties of the fitted parameters are not given, but we discuss the effects of parameter variations on our results in the following section.

    \begin{table}[ht!]
        \caption{\label{tab:modelparameters}Jet model parameters for Cygnus X1. Values taken from \mbox{scenario A}
        of \citet{Pepe2015}.}
        \centering
        \begin{tabular}{llrl}
        \hline\hline
        Parameter & Symbol & Value & Units\\
        \hline
        Jet half-opening angle & $\theta_{\mathrm{jet}}$ & $2$ & ${\mathrm{deg}}$\\
        Jet launching distance & $z_{0}$ & $1.1\times10^{8}$ & $\mathrm{cm}$\\
        Base of acceleration region & $z_{\mathrm{min}}$ & $2.8\times10^{8}$ & $\mathrm{cm}$\\
        Top of acceleration region & $z_{\mathrm{max}}$ & $1.9\times10^{12} $ & $\mathrm{cm}$\\
        Jet bulk Lorentz factor & $\Gamma$ & $1.25$ &\\
        Magnetic power-law index & $\alpha$ & $1.9$ &\\
    Jet luminosity & $L_{\mathrm{jet}}$ & $10^{38}$ & $\mathrm{erg\,s^{-1}}$ \\
        Relativistic power fraction & $q_{\mathrm{rel}}$ & $0.1$ &\\
        p-e luminosity ratio & $a$ & $39$ &\\
        Injection spectral index & $p$ & $2.4$ &\\
        Minimum particle energy & $E_{min}$ & $95.4$ & $mc^{2}$\\
        Acceleration efficiency & $\eta$ & $6\times10^{-4}$ &\\
        \hline
        \end{tabular}
        \end{table}

\begin{figure}
    \centering
    \includegraphics[width=0.95\hsize]{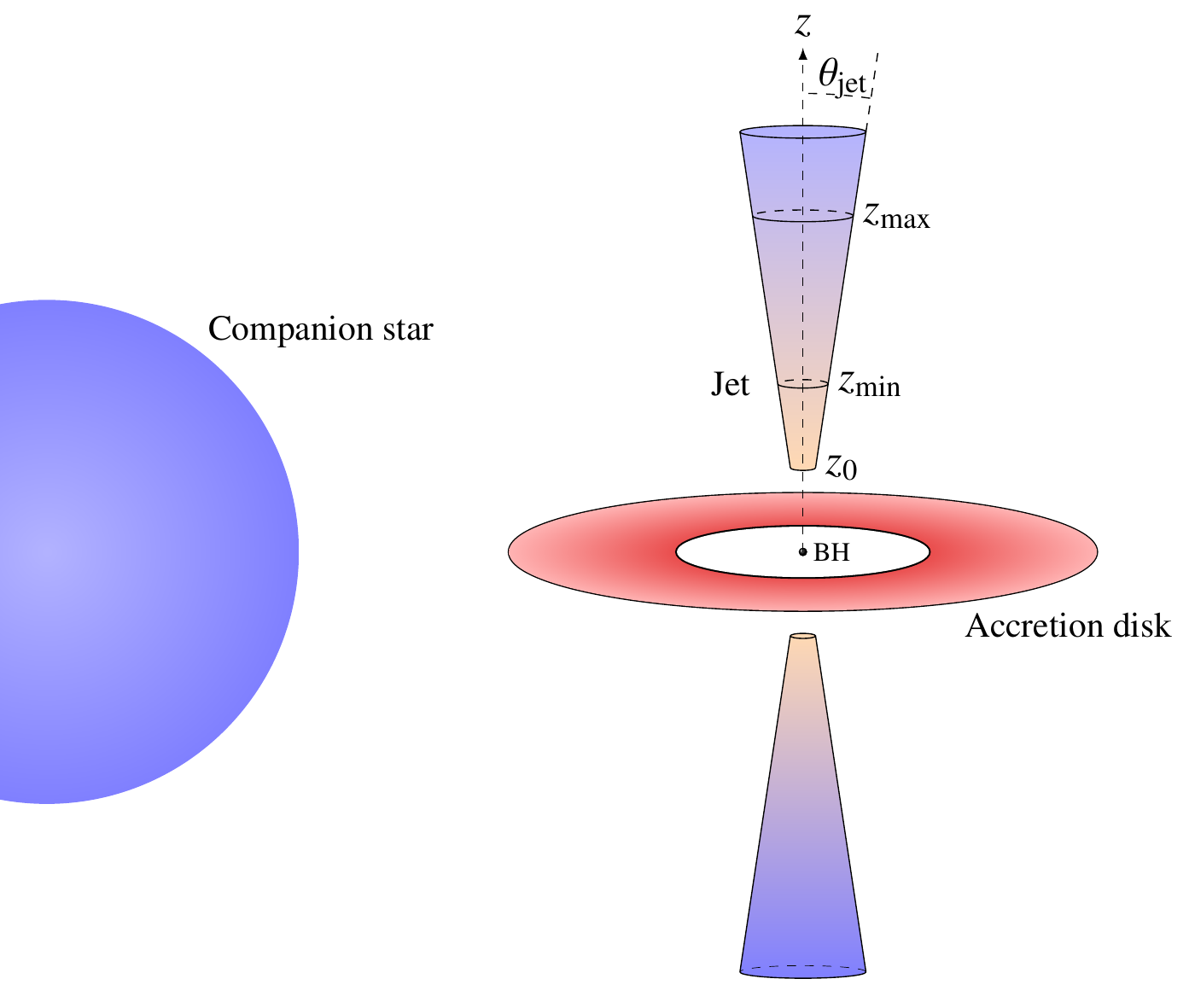}
    \includegraphics[width=0.75\hsize]{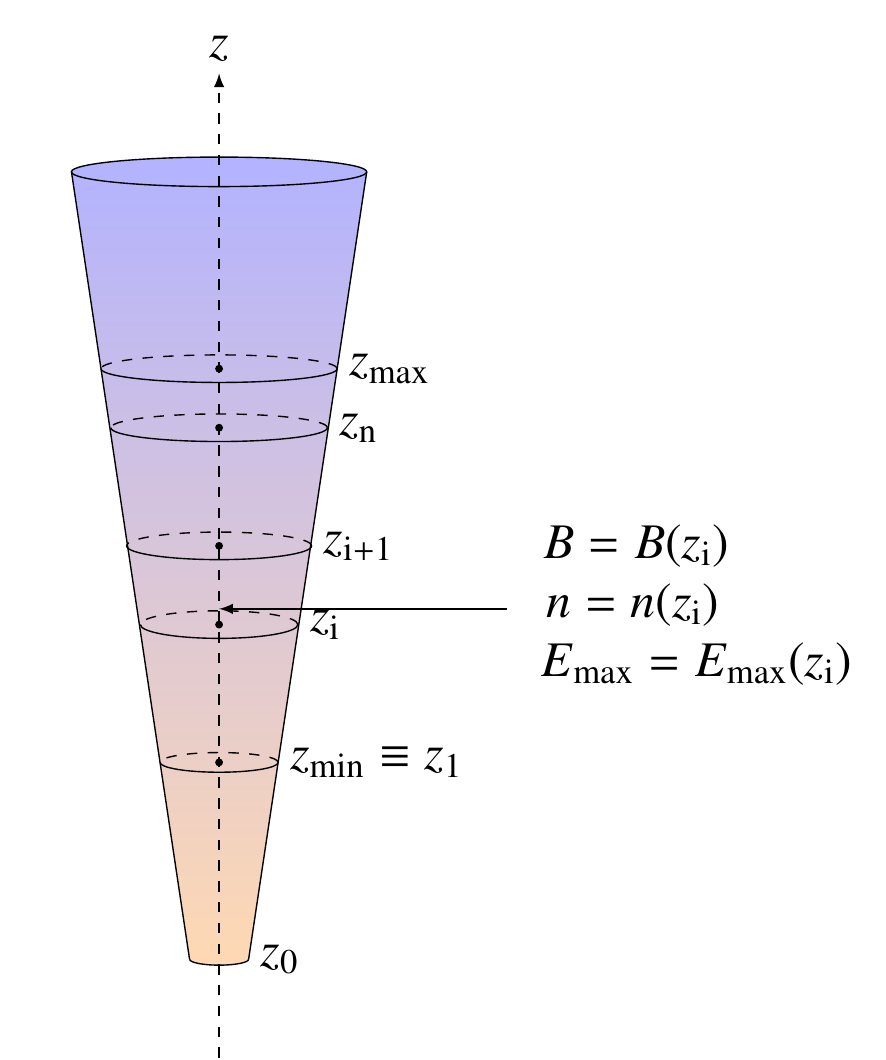}
    \caption{\emph{Top}: Schematic picture of a microquasar (not to scale). The main parameters used in the model are indicated. \emph{Bottom}: Discretisation mesh scheme defining a set of $n$ logarithmically spaced points $z_{\mathrm{i}}$.}
    \label{Fig:microquasar}
    \end{figure}

   Figure~\ref{Fig:microquasar} presents a schematic picture of a MQ, showing the jets launched from the vicinity of the compact object. We adopt a lepto-hadronic, conic jet model with semi-aperture angle $\theta_{\mathrm{jet}}$. The flow propagates along the $z$ axis with a macroscopic (bulk) Lorentz factor $\Gamma$, and carries a total power $L_{\mathrm{jet}}$. The jet is pervaded by a magnetic field $B(z) = B_{0}\,(z_{0}/z)^{\alpha}$, where 
   $\alpha$ is the magnetic index, and $z_{0}$ and $B_0$ are the position and magnetic field of the jet-launching region, respectively. The latter is derived assuming equipartition between magnetic and kinetic energy at the jet base,
   
   \begin{equation}
   \frac{B_{0}^{2}}{8\pi} = \frac{L_\mathrm{jet}}{\pi r_{0}^{2}v_{\mathrm{jet}}},
   \label{magneticfield}
   \end{equation}
   
   \noindent where $r_0$ is the jet radius at its base, and $v_{\mathrm{jet}}$ the jet bulk
   velocity. Hereafter we work in the jet frame, where the bulk is at rest.
   
   We assume that internal shocks in the jet accelerate protons to relativistic energies in a region confined between $z_{\mathrm{min}}$ and $z_{\mathrm{max}}$, and that cooling of these particles takes place locally. A fraction $q_{\mathrm{rel}}$ of the jet power is deposited in relativistic particles, with a proton-to-electron luminosity ratio $a$ (i.e. $a = L_{\mathrm{p}} / L_{\mathrm{e}}$),
   
   \begin{equation}
       q_\mathrm{rel} L_\mathrm{jet} = (1 + a^{-1}) L_\mathrm{p}, 
   \end{equation}
   
   \noindent where $L_{\mathrm{p}}$ is the total power in the relativistic proton population. Injection of protons in the acceleration region is described by a source rate density
   
   \begin{equation}
   \centering
        Q(E_\mathrm{p},z) = Q_0 \, (E_\mathrm{p})^{-p} \, 
        \exp [- E_\mathrm{p} / E_\mathrm{p,max}(z)],
   \end{equation}
   
   \noindent where $Q$ gives the number of protons injected per unit time and volume, and per proton energy interval $(E_{\mathrm{p}}, E_{\mathrm{p}} + \mathrm{d}E_{\mathrm{p}})$. The maximum attainable energy is $E_{\mathrm{p,max}}$, \textbf{$p$} is the injection spectral index, and $Q_0$ is a normalisation constant obtained from 
   
   \begin{equation}
   \centering
        L_\mathrm{p} = \int_{V} \int_{E_p} Q(E_\mathrm{p},z) \, 
        \mathrm{d}E_\mathrm{p} \, \mathrm{d}V,
   \end{equation}
   
   \noindent with $V$ the volume of the acceleration region.
   
   Neutrons are produced in inelastic collisions of relativistic protons with baryons in the bulk. We disregard the inverse process, that is, the production of protons by relativistic neutrons colliding with bulk protons. Below, we show that the rate of this process is negligible compared to the neutron escape rate. Photomeson production takes place beyond the threshold $\epsilon' \approx 145~\text{MeV}$, where $\epsilon'$ is the target photon energy in the frame of the relativistic proton. This corresponds to proton energies of $\sim 1~\text{PeV}$ for an X-ray photon field (e.g. that of the accretion disc or a corona). As we see below, contrary to AGN jets, these energy values are barely achieved by protons in a typical MQ jet. Other potential fields to be considered have lower-energy photons (e.g. synchrotron, stellar radiation) and therefore proton energies must rise above $1~\text{PeV}$ in order for the process to take place. 
        Using a very similar model, \citet[see their Fig. 5]{Pepe2015} found that the cooling rate for $p$-$\gamma$ interaction against X-ray photons coming from a hot corona is several orders of magnitude lower than that for $p$-$p$ collisions. Therefore, the production of neutrons through the former process is not taken into account.
        
        In a stationary regime, the relativistic proton and neutron populations obey the following transport equations:
        
        \begin{eqnarray}
     \label{eq:transportp}
     \frac{\partial}{\partial E_\mathrm{p}} [b_\mathrm{p} N_\mathrm{p}] = Q - 
     \Lambda - t^{-1}_{\mathrm{esc}}\,N_\mathrm{p}, \\
     \label{eq:transportn}
     0 = Q_\mathrm{n}  - 
     t^{-1}_{\mathrm{esc,n}}N_\mathrm{n}. 
        \end{eqnarray}
        
        \noindent
        In these equations, $b_{\mathrm{p}}$ is the total energy loss rate of protons through all cooling processes (adiabatic plus radiative), and $N_{\mathrm{p(n)}}$ is the proton (neutron) spectral density. The term $Q_{\mathrm{n}}$ represents a source of neutrons due to proton--proton collisions, which is related to the proton sink term $\Lambda$ through $Q_\mathrm{n}(E_\mathrm{n}) = \Lambda(2 E_\mathrm{n})$, as roughly half of the proton energy in a $p$-$p$ collision is transferred to the neutron. The total neutron spectral power is then computed as
        
        \begin{eqnarray}
    P_{\mathrm{n}}(E_{\mathrm{n}}) \equiv \frac{\mathrm{d}E}{\mathrm{d}E_{\mathrm{n}}\mathrm{d}t} = \int_{V}\,E_{n}\,Q_{\mathrm{n}}(E_{n})\,\mathrm{d}V,
    \label{eq:neutronpower}
    \end{eqnarray}
        
        \noindent where [$P_{\mathrm{n}}$] = $\mathrm{s^{-1}}$. Neutrons experience neither adiabatic nor radiative cooling.

    The last term of  Eqs.~\ref{eq:transportp}--\ref{eq:transportn} represents the escape of particles; following \citet{Romero&Vila2008} we adopt
        $t^{-1}_{\mathrm{esc}} \sim v_{\mathrm{jet}} (z_\mathrm{max} - z_\mathrm{min})^{-1}$ as a characteristic escape rate for charged particles. Neutrons are not advected by the plasma,  and therefore their escape rate is $t^{-1}_{\mathrm{esc,n}} \sim c / l$, with $c$ the speed of light and $l$ a characteristic size of the jet. A rather conservative lower limit is $t^{-1}_{\mathrm{esc,n}} > c (z_\mathrm{max} - z_\mathrm{min})^{-1}$, which arises from neutrons travelling through the whole acceleration region and escaping the jet through its head. Indeed, as neutron production is isotropic in the jet frame, most neutrons will escape by the side of the jet, increasing the rate by a factor of the order of $\theta_\mathrm{jet}^{-1}$. As we show below, our key results do not depend on the exact value of $t^{-1}_{\mathrm{esc,n}}$.
        
        Neutron decay is not considered at this stage because its rate is negligible with respect to escape. For our fiducial model $z_\mathrm{max} - z_\mathrm{min} \approx 2 \times 10^{12}\,\mathrm{cm}$, and we obtain $t^{-1}_{\mathrm{esc,n}} \gtrsim 1.5 \times 10^{-2}~\mathrm{s}^{-1}$. The neutron decay rate is $t^{-1}_{\mathrm{d}} = \gamma_{\mathrm{n}}^{-1}\,\tau_{\mathrm{d}}^{-1}$, where $\gamma_{\mathrm{n}}$ is the neutron Lorentz factor and $\tau_{\mathrm{d}}^{-1} = 1.13\times 10^{-3}\,\mathrm{s}$ is the inverse of the neutron lifetime. Therefore, neutron decay within the jet can be neglected in our model for any value of $\gamma_{\mathrm{n}}$. This preserves the locality of the model, because neutron decay inside the jet would couple the populations of protons in different regions.

    Radiative cooling processes considered for protons
        are synchrotron radiation due to the motion in the jet magnetic field and $p$-$p$ interactions leading to $\gamma$-ray emission through pion decay.
        The synchrotron cooling rate $t_\mathrm{syn}^{-1}$ is taken from \citet{Blumenthal1970}. For proton--proton inelastic scattering, we consider the two main branches 
        
        \begin{eqnarray}
        \label{eq:NeutronChannel}
     p + p \longrightarrow p + n + \pi^{+} + l\pi^{0} + k(\pi^{+} + \pi^{-}), \\
     \label{eq:ProtonChannel}
     p + p \longrightarrow p + p + l\pi^{0} + k(\pi^{+} + \pi^{-}),
        \end{eqnarray}
        
        \noindent where $l$ and $k$ are the neutral and charged pion multiplicities, respectively. The total cooling rate of the process is given by 
        
        \begin{equation}
        t^{-1}_\mathrm{pp} = \sigma_\mathrm{pp} \, n_{\mathrm{b}} \, v_\mathrm{p} \, K_\mathrm{pp}, 
   \end{equation}
        
        \noindent where $K_\mathrm{pp}$ and $\sigma_\mathrm{pp}$ are the $p$-$p$ inelasticity and cross-section, 
        respectively, $v_\mathrm{p}$ is the velocity of relativistic protons relative to that of bulk ones (we adopt $v_{\mathrm{p}} \approx c$), and \mbox{$n_{\mathrm{b}}\approx L_{\mathrm{jet}} (1-q_{\mathrm{rel}}) [m_{\mathrm{p}}c^{2} v_{\mathrm{jet}} \pi \tan^{2}\theta_{\mathrm{jet}}\,z^{2}]^{-1}$}
        the bulk proton density in the reference frame of the jet \citep[cf.][]{VilaRomeroCasco2012}. We compute $K_\mathrm{pp}$ and $\sigma_\mathrm{pp}$  following \citet{Kafexhiu2014}. A parameterisation of the inclusive cross section for the channel of Eq.~(\ref{eq:NeutronChannel}) at low energies is also given by these authors. For consistency with that parameterisation and with measurements of the inclusive cross-section at higher energies \citep[e.g.][]{Engler1975, Flauger&Monnig1976, Adare2013, Adriani2018}, we assume a probability of $0.16$ as a low, conservative value for the production of a neutron in a proton--proton collision, that is, $\Lambda(E_{\mathrm{p}}) = 0.16\,N_\mathrm{p}\,t_\mathrm{pp}^{-1}$. The proton energy loss rate given by the interaction channel (\ref{eq:ProtonChannel}) is therefore
        
        \begin{equation}
        b_{\mathrm{pp}} := \frac{\mathrm{d}E_{\mathrm{p}}}{\mathrm{d}t} = - 0.84\,\sigma_\mathrm{pp} \, n_{\mathrm{b}} \, v_\mathrm{p} \, K_\mathrm{pp} \, E_{\mathrm{p}}.
        \end{equation}

The neutron--proton collision rate is $t_{\mathrm{np}}^{-1} \sim t_{\mathrm{pp}}^{-1}$; therefore, neutrons will escape without interacting with bulk protons if $t_{\mathrm{pp}}^{-1} \ll t_{\mathrm{esc,n}}^{-1}$. As the opening angle is small, most neutrons will escape through the side of the jet, for which $t_{\mathrm{esc,n}}^{-1} \approx c / r(z) \approx 3\times 10^{3}~(z_{\mathrm{min}} / z)~\mathrm{s}^{-1}$, and the condition for escaping without interacting is fulfilled all along the region of interest.

        Protons are also cooled through adiabatic losses, at a rate given by 
        
        \begin{equation}
        \centering
        t^{-1}_{\mathrm{ad}} = \frac{2}{3} \frac{v_{\mathrm{jet}}}{z}. 
   \end{equation}

Densities, cooling, source, and sink terms in the transport equations depend on particle energies and $z$. As there are no explicit spatial derivatives, Eqs.~(\ref{eq:transportp})--(\ref{eq:transportn}) become a set of coupled ordinary differential equations at each point $z$ along the jet axis. To solve them, 
we discretise the functions in a mesh along the $z$ axis in the region of interest. For each point in the mesh, we solve the system of equations numerically via the Picard method, using the one-zone approximation (i.e. assuming that what happens in one region has no effect on any of the others) and explicit differences to compute energy derivatives. Because the jet extends over several orders of magnitude in $z$, we adopt a logarithmic mesh (see Fig.~\ref{Fig:microquasar}). We use standard quadrature methods to compute the volume integrals required to obtain the properties of the whole acceleration region. 

        The boundary condition for solving the transport equations is $N_{\mathrm{p}}(E_{\mathrm{p, max}}) = 0$. The maximum energy of protons is obtained from the condition $t^{-1}_{\mathrm{acc}}(E_{\mathrm{p, max}}) = t^{-1}_{\mathrm{loss}}(E_{\mathrm{p, max}})$, where the total loss rate is $t^{-1}_{\mathrm{loss}} = t^{-1}_\mathrm{ad} + t^{-1}_\mathrm{sync} + t^{-1}_\mathrm{pp}$, and 
        $t^{-1}_{\mathrm{acc}}$ is the proton acceleration rate.
                We assume that a diffusive shock mechanism operates to accelerate charged protons at a rate of 
         
        \begin{equation}
        \centering
        t^{-1}_{\mathrm{acc}} = \frac{\eta\,e\,c\,B}{E_i} 
   ,\end{equation}
        
\noindent where $e$ is the elementary charge, and $0 \leq \eta \leq 1$ an efficiency parameter.

\section{Neutron production and escape}
\label{sec:neutron_production}
\subsection{The jet of \object{Cygnus X-1}}     
\label{sec:CygnusX1}

        \begin{figure*}
    \centering
    \includegraphics[width=0.98\hsize]{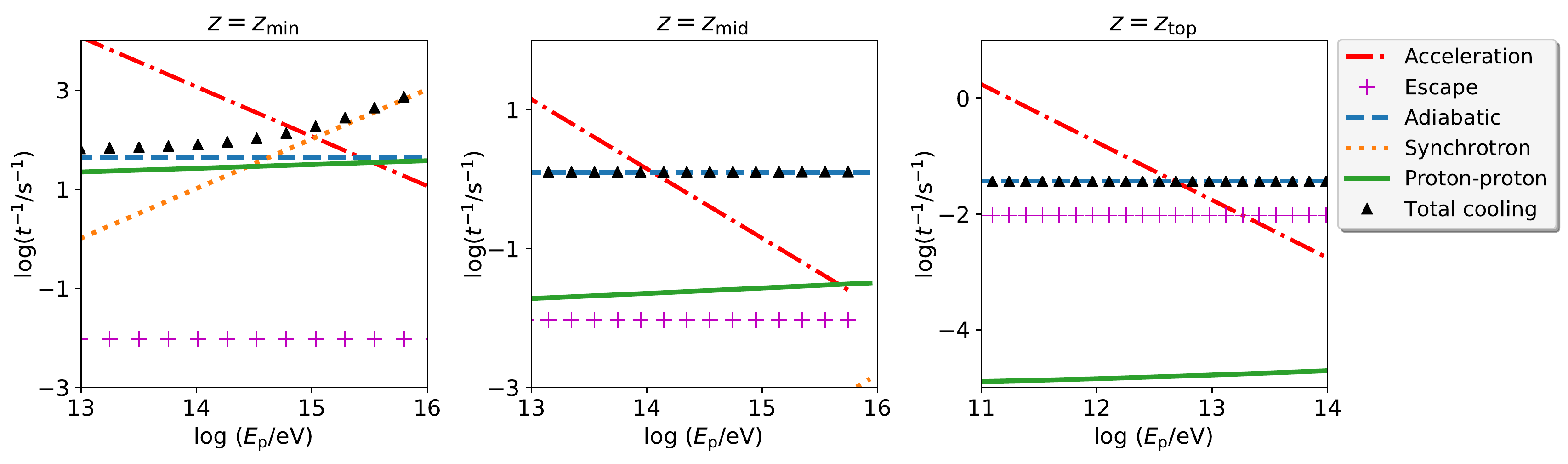}
    \includegraphics[width=0.98\hsize]{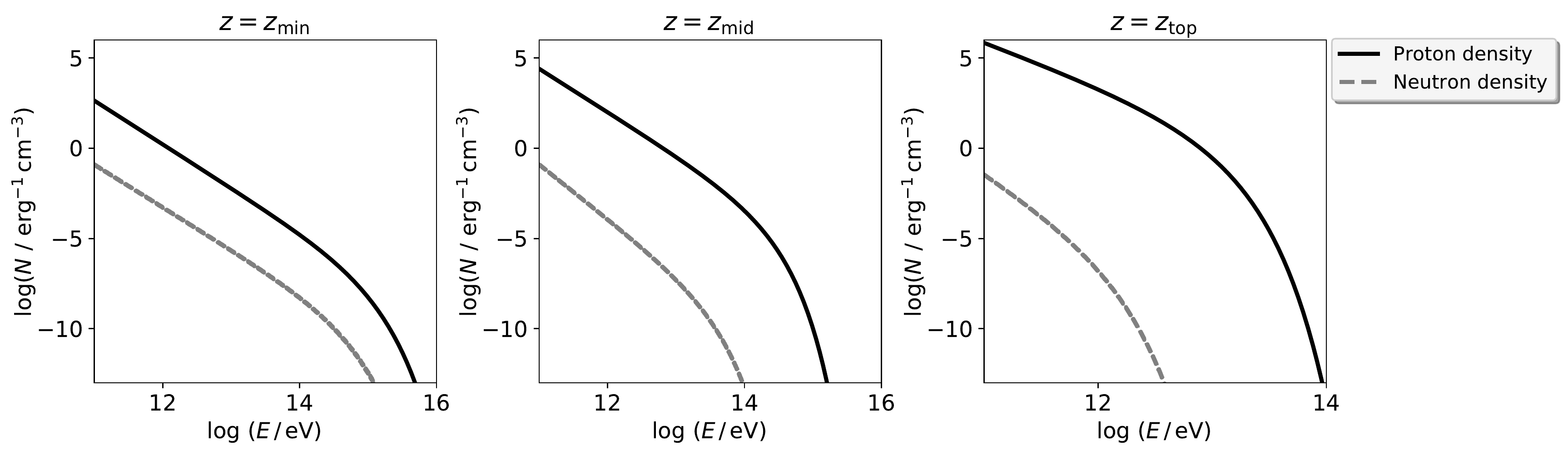}
    \caption{Energetics of hadrons at the base ($z_{\mathrm{min}}$, left panels), logarithmic midpoint ($z_{\mathrm{mid}}$, middle panels), and top ($z_{\mathrm{top}}$, right panels) of the acceleration region for the Cygnus X-1 model. \emph{Top:} Cooling and acceleration rates for protons. The plots show the loss rates for proton--proton (green, solid line), synchrotron (yellow, dotted line), escape (magenta, + symbols), and adiabatic losses (light blue, dashed line), the total loss rate (black triangles), and the acceleration rate (red, dash-dotted line). \emph{Bottom:} Proton (black, solid line) and neutron (grey, dashed line) densities for the same regions as the top plots.
        In all cases, adiabatic losses are dominant and the proton population is many orders of magnitude denser than the neutron one.}
    \label{Fig:cyg_rates}
    \end{figure*}

    Figure~\ref{Fig:cyg_rates} shows the proton acceleration and cooling rates for our fiducial model, at the base, middle, and top of the acceleration region, together with the neutron and proton densities at the same places.
        At the base, protons reach maximum energies of about $10^{15}\,\mathrm{eV}$, which is also an upper limit for the energies of neutrons, because the latter are about half of the former. At higher $z,$ this value decreases because the acceleration rate, governed by the magnetic field, varies as $z^{-1.9}$, whereas the total loss rate changes as $\sim z^{-1}$. Adiabatic losses dominate along the
    whole jet. We therefore expect that the most energetic neutrons are mainly produced in regions near the base of the jet. On the other hand, we observe that the proton density is at least two orders of magnitude higher than the neutron density. The difference increases with the distance to the base of the jet. This is due to the fact that the density of target protons decreases with $z$. Thus, the proton spectrum is roughly the same as it would be without considering neutron production (we recall that the latter process is the unique sink for the proton population besides the escape), and the same applies to the gamma-ray SED produced by hadrons. This implies that, at least with the sensitivity of present instruments, there is no possibility of detecting the hadronic nature of jets by any signature produced by neutrons in their SEDs.
    
    \begin{figure}
    \centering
    \includegraphics[width=0.95\hsize]{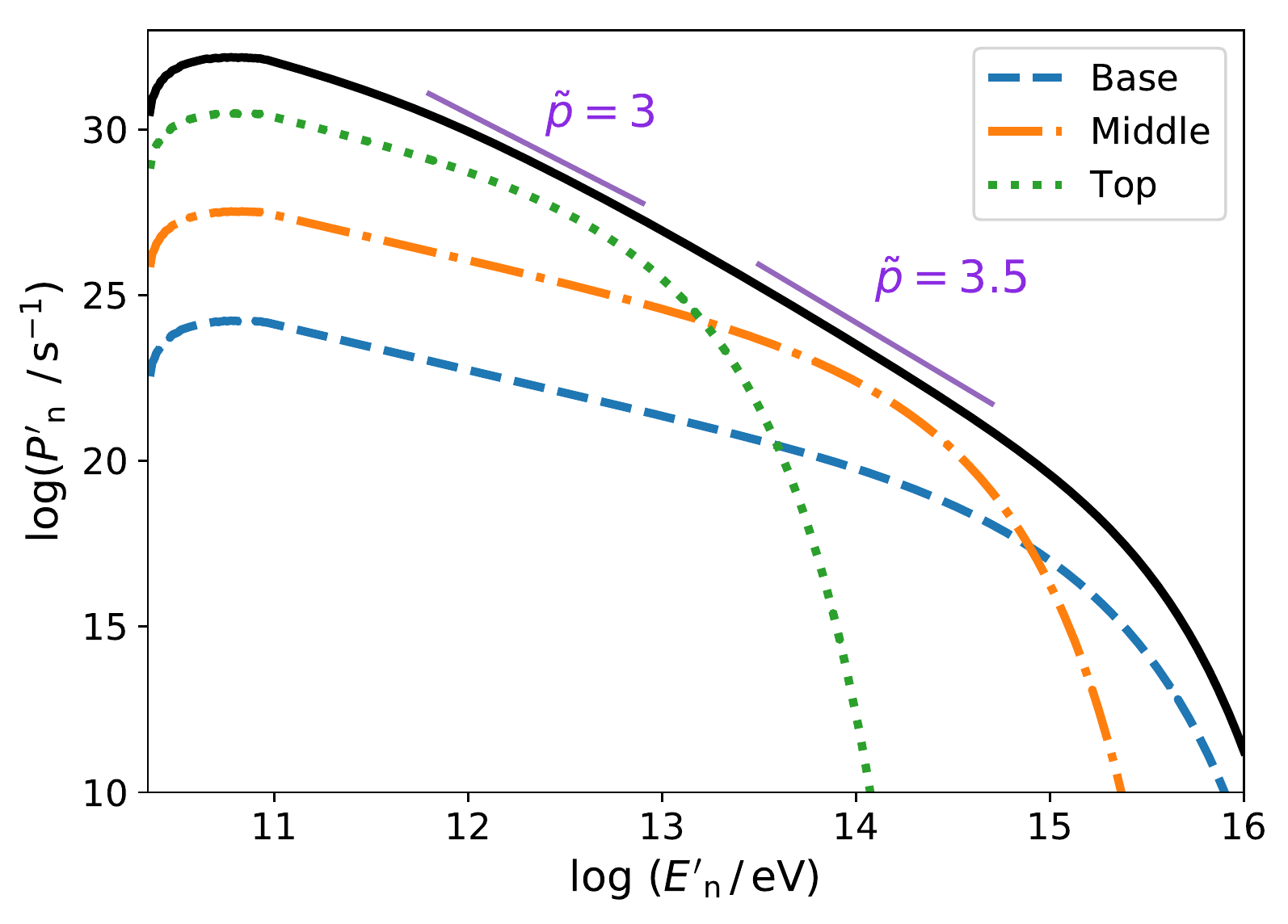}
    \caption{Spectral power of neutrons produced in collisions between relativistic protons with bulk ones computed in the ISM reference frame for the Cygnus X-1 model. We show the contribution of 
                 base, middle, and top regions in the jet (dashed, dash-dotted and dotted lines, respectively).
                 The black solid line accounts for the total power. Violet segments represent power-law distributions with spectral index $\Tilde{p}$.
             }
    \label{fig:neutronpower}
    \end{figure}
    
    The steady-state neutron density depends on the escape rate (see Eq.~\ref{eq:transportn}), and represents the population of neutrons in transit before escaping. However, the injection rate of neutrons in the ISM is independent of the escape regime, because it is determined by the neutron production rate alone. Assuming that the neutron distribution is isotropic in the reference frame of the jet, the power injected in neutrons into the ISM, in the reference frame of the latter, will depend on the $z$-axis direction cosine $\mu$ in our model. 
    To compute the spectral power of the neutron population injected into the ISM, we transform Eq.~(\ref{eq:neutronpower}) according to 
    
    \begin{eqnarray}
    \label{totalpower}
    P'_{\mathrm{n}}(E'_{\mathrm{n}}) = \frac{1}{2}\int_{-1}^{1} P_{\mathrm{n}}(E_{\mathrm{n}}) \frac{\mathrm{d}\mu}{\mathrm{d}\mu'} \frac{\mathrm{d}E_{\mathrm{n}}}{\mathrm{d}E'_{\mathrm{n}}}\mathrm{d}\mu',
    \end{eqnarray}  
    
    \noindent where
    
    \begin{eqnarray}
    \label{}
    \frac{\mathrm{d}\mu}{\mathrm{d}\mu'} = \frac{(1 + \beta_{\mathrm{jet}}\mu')^{2}}{1-\beta_{\mathrm{jet}}^{2}},\\
    \frac{\mathrm{d}E_{\mathrm{n}}}{\mathrm{d}E'_{\mathrm{n}}} = \Gamma (1-\beta_{\mathrm{jet}}\mu'),
    \end{eqnarray}
    
    \noindent where $\beta_{\mathrm{jet}} = v_{\mathrm{jet}}/c$. In Eq.~(\ref{totalpower}), the factor $1/2$ comes from a previous integration in azimuth, and non-primed quantities refer to the rest frame of the jet. For simplicity, hereafter we use primed quantities for variables measured in the reference frame of the ISM. 
    
    Figure~\ref{fig:neutronpower} shows the spectral power of the neutron population that escapes from the jet for the \object{Cygnus X-1} case, which is computed in the reference frame of the ISM. The contributions of different regions of the jet are shown in the same way as in Fig.~\ref{Fig:cyg_rates}. The population produced at the base of the jet contributes to the high-energy spectral region, while that coming from higher zones dominates the low-energy region. The minimum energy is the same in all cases ($E_\mathrm{n,min} \approx 0.5 E_{\mathrm{p,min}}$) because that of protons is an input parameter, which for the fiducial model is the value determined by \citet{Pepe2015} via SED fitting. Regarding the total production, most of the power is injected in low-energy neutrons. The population presents a spectral index ($\Tilde{p}$) of $\approx 3$ at neutron energies of $E_{\mathrm{n}} \approx 10^{11-12}~$eV, and steepens at higher energies, where the values of the spectral index shift up to $\Tilde{p} \approx 3.5$ at $E_{\mathrm{n}} \approx 10^{14-15}~$eV. The slope of the curve depends on the proton loss process that dominates at each energy. We note from Fig.~\ref{Fig:cyg_rates} that, at each region of the jet, the relative contributions between adiabatic, proton--proton, and escape losses are modified. For the same reason, the maximum proton energies are also different in each region. Both effects contribute to a variation of the spectral index of the proton population, which is reflected in the spectral index of the neutron population, as shown in Fig.~\ref{fig:neutronpower}. The complete neutron population carries a total power of $\approx 3.3 \times 10^{31}~\text{erg}~\text{s}^{-1}$, which is of the order of $10^{-7}L_{\mathrm{jet}}$.
    
\subsubsection{Stellar wind contribution}

The wind of the companion star may penetrate the jet and mix with its matter, thereby increasing its density and enhancing proton--proton interactions \citep{Romero2003}. The stellar-wind proton number density is 
        given by
        
        \begin{eqnarray}
    n_{\mathrm{w}} = \frac{\dot{M}}{4\pi r^{2}v_{\mathrm{w}} m_{\mathrm{H}}},
    \end{eqnarray}  
    
    \noindent where $\dot{M}$ is the mass-loss rate, $v_\mathrm{w}$ is the velocity of the wind, and $m_\mathrm{H}$ is the mass of the hydrogen atom. The standard velocity profile for a line-driven wind is given by \citep{Lamers1999}
    
    \begin{eqnarray}
    v_{\mathrm{w}} = v_{\infty} \left( 1 - \frac{R_{\star}}{r} \right)^{\delta},
    \end{eqnarray}  
    
    \noindent where $v_{\infty}$ is the terminal velocity, $R_{\star}$ the radius of the star, and \mbox{$0.5 \leq \delta \leq 1$}.
    
For our fiducial model we adopt $R_{\star} \approx 20 R_{\sun}$, $v_{\infty} \approx 2\,100 ~\mathrm{km}~ \mathrm{s}^{-1}$ and $\dot{M}\approx 3 \times 10^{-6} ~\mathrm{M}_{\odot} ~\mathrm{yr}^{-1}$ \citep[e.g.][]{Herrero1995, Yan2008}, and the binary system separation $a_{*} \approx 3\times 10^{12}\,\mathrm{cm}$ \citep[e.g.][]{Iorio2008}. Assuming that all the material of the wind mixes with the jet, the bulk-to-wind density ratio is $\approx 8 \times 10^{-2}[1 + (a_{\star} / z)^{2}]$. Thus, at the base of the acceleration region the bulk density overcomes that of the wind ($n_{\mathrm{p}}/n_{\mathrm{w}}\approx 10^{6}$), while near the top of the region the wind density becomes significant ($n_{\mathrm{p}}/n_{\mathrm{w}}\approx 10^{-1}$). The contribution to the total neutron power is $\approx 6 \times 10^{31}~\text{erg}~\text{s}^{-1}$, and is roughly the same for any value of $\delta$ in the given range. This contribution is twice that of the neutron production with bulk protons as targets. 
We note that this wind scenario is extreme in the sense that the star is very close to the jet, and has a very high mass-loss rate. Varying the companion properties would then decrease the contribution of the wind material to neutron production. To achieve
higher neutron luminosities, we therefore explore scenarios with different jet parameters.
        
\subsection{Other jet scenarios}
        
\begin{figure*}
    \centering
    \includegraphics[width=0.98\hsize]{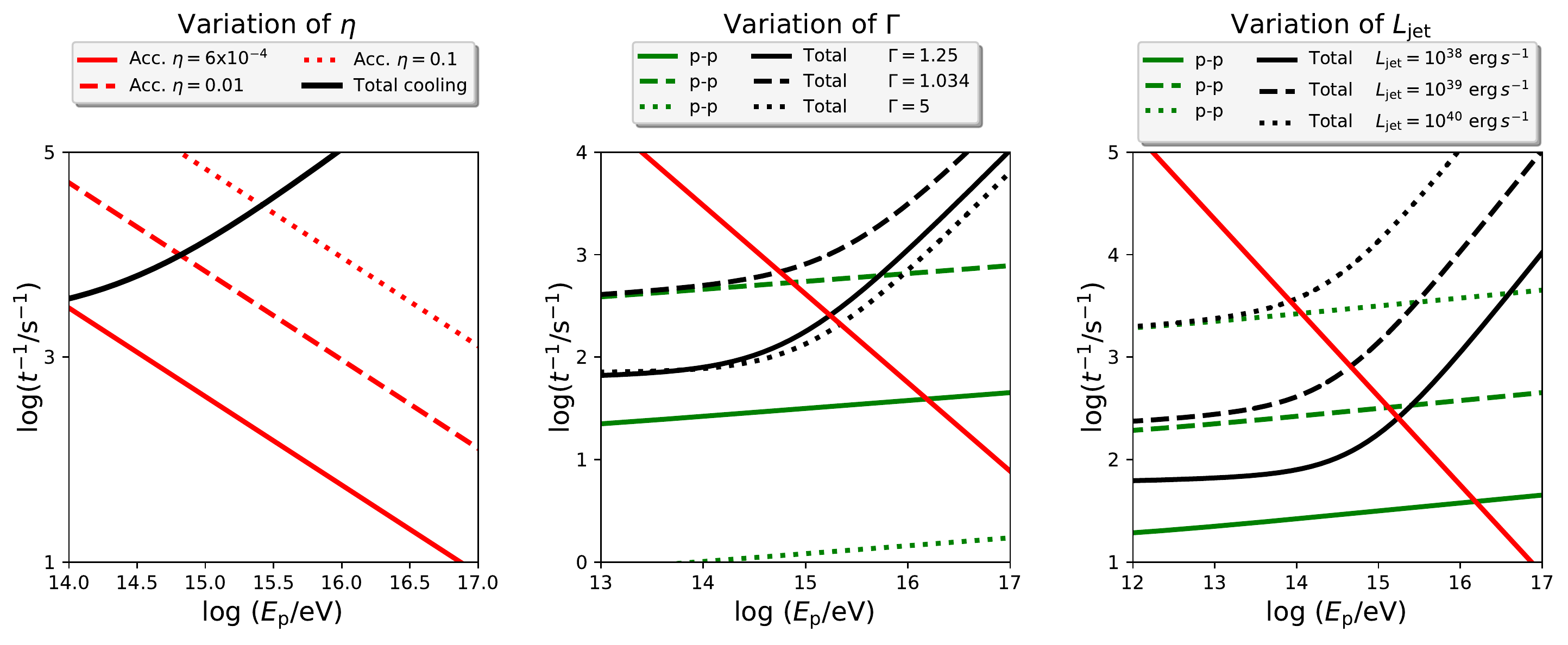}
    \caption{\emph{Left:} Total loss-rate (black solid line) and acceleration rates (red lines) for different values of the efficiency parameter $\eta$: the Cygnus X-1 model (solid line, $\eta = 6\times 10^{-4}$), $\eta = 10^{-2}$ (dashed line) and $\eta=0.1$ (dotted line). \emph{Middle:} Acceleration rate (red solid line, $\eta = 6\times 10^{-4}$), total loss rate (black lines), and proton--proton loss-rate (green lines) for different bulk Lorentz factors: that of Cygnus X-1 (solid line, $\Gamma = 1.25$), $\Gamma = 1.034$ (dashed line), and $\Gamma = 5$ (dotted line). \emph{Right:} Acceleration rate (red solid line) for the case of Cygnus X-1, and total loss rate (black lines) and proton--proton loss-rate (green lines) for different jet luminosities, i.e. $L_{\mathrm{jet}}$: that of Cygnus X-1 (solid line, $L_{\mathrm{jet}} = 10^{38}~\mathrm{erg~s}^{-1}$), $L_{\mathrm{jet}} = 10^{39}~\mathrm{erg~s}^{-1}$ (dashed line), and $L_{\mathrm{jet}} = 10^{40}~\mathrm{erg~s}^{-1}$ (dotted line). All panels refer to the jet base.}
    \label{Fig:rates_models}
    \end{figure*}

        We have shown that, although a significative number of neutrons are indeed produced in a typical MQ jet, the energy carried by them to the ISM is small in the considered case. In this section we perform a variation of the main parameters of the jet model: the bulk Lorentz factor $\Gamma$, the efficiency of the acceleration $\eta$, the spectral index $p$, and the jet luminosity $L_{\mathrm{jet}}$. The rest of the parameters remain those of the \object{Cygnus X-1} model. In particular, we fix the value of the magnetic index ($\alpha = 1.9$). The magnetic field is expected to have a dominant poloidal component near the jet base that becomes toroidal towards higher distances. This would result in a variation of the magnetic index from $\alpha \approx 2$ to $\alpha \approx 1$. However, as we see from Fig.~\ref{Fig:cyg_rates}, synchrotron losses dominate close to maximum energies and just near the jet base. 
         In this region, the magnetic field is near the equipartition value, and therefore synchrotron may play a role in limiting the maximum neutron energy at the jet base only for jets with large kinetic luminosities. 
        Beyond $z\sim z_{\mathrm{mid}}$, the synchrotron rate 
        is negligible regardless of the value of the magnetic index. Figure~\ref{Fig:rates_models} shows how the energetics of the hadron populations and the neutron production are modified in different scenarios. As we see, a more effective acceleration shifts the maximum energy of protons by $\gtrsim 1$ order of magnitude. On the other hand, lower bulk Lorentz factors lead to an increase in the proton--proton interaction rate, and hence the neutron production rate. For this parameter, we used alternative values, namely those of the MQ with the lowest jet velocity measured  \citep[$\Gamma = 1.034$, for \object{SS 433};][]{Chaty2007} and the value adopted by \citet[$\Gamma = 5$]{Heinz&Sunyaev2002}. Finally, we observe that increasing the jet luminosity increases both the relativistic and thermal proton densities, in turn increasing the neutron production rate.
        
\begin{figure*}
    \centering
    \includegraphics[width=0.98\hsize]{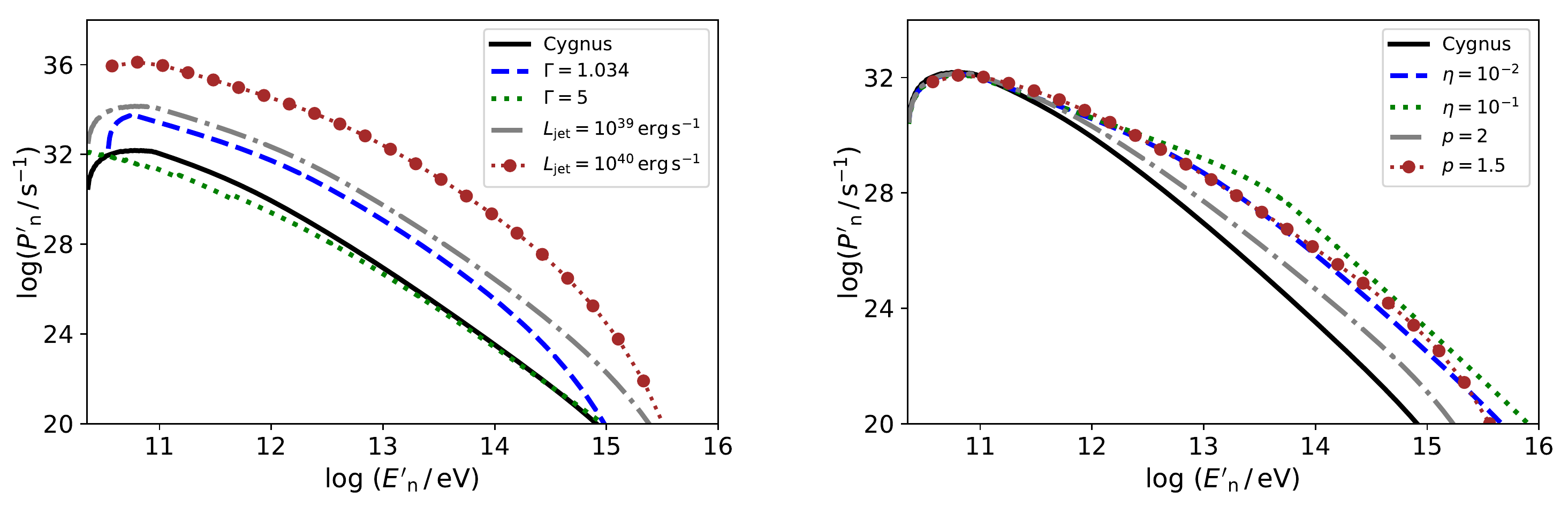}
    \caption{Spectral power of relativistic neutron population for different model parameters, in comparison to that of \object{Cygnus X-1}. In all cases, the spectra are computed in the ISM reference frame. \emph{Left}: Variation of macroscopic parameters $\Gamma$ and $L_{\mathrm{jet}}$. \emph{Right}: Variation of microscopic parameters $\eta$ and $p$. }
    \label{Fig:spectralpower_models}
    \end{figure*}

        We show the spectral power of neutrons for eight models in Fig.~\ref{Fig:spectralpower_models}. The variation of microscopic parameters such as $\eta$ and $p$ changes the hardness of the neutron population, increasing it as the proton injection becomes harder or the acceleration more efficient. These parameters produce minor variations in the total neutron power. The minimum energy of the relativistic protons is a parameter given in the rest frame of the jet; it changes the way in which the total energy input is distributed and is related to the value at which the neutron population peaks in the ISM frame. However, it does not have an impact on the spectral index or the shape of the distribution in general. On the other hand, macroscopic parameters do not significantly modify the spectral index, but do affect the general energetics of the population. A decrease in the bulk Lorentz factor increases the total power, while approximately preserving the shape of the spectrum. Higher Lorentz factors do not lead to significant changes in the spectrum because the lower densities of target protons limit the energy loss by the $p$-$p$ channel.  
        The neutron spectral power is also highly dependent on the jet luminosity, because an increase in the latter increases both the proton population energy and collision rate. For the eight models explored, we computed the total power in relativistic neutrons, $L_{\mathrm{n}}$. The results are summarised in Table \ref{Tab:models}.
    
    \begin{table}[ht!]
        \caption{\label{Tab:models}Total power in relativistic neutron population.}
        \centering
        \begin{tabular}{cccccc}
        \hline\hline
        \!\!\!Model & \!\!\!$p$ & \!\!\!$\Gamma$ & \!\!\!$\eta$ & \!\!\!\!$L_{\mathrm{jet}}~[\mathrm{erg~s^{-1}}]$ & \!\!\!\!$L_{\mathrm{n}}$ [$\mathrm{erg~s^{-1}}$] \\
        \hline
        \!\!\!\object{Cygnus X-1} & \!\!\!$2.4$ & \!\!\!$1.25$ & $6\times 10^{-4}$ & \!\!\!\!$10^{38}$ & \!\!\!\!$3.3 \times 10^{31}$\\
        \!\!\!1 & \!\!\!$2.4$ & \!\!\!$1.034$ & $6\times 10^{-4}$ & \!\!\!\!$10^{38}$ & \!\!\!\!$1.0 \times 10^{33}$\\
        \!\!\!2 & \!\!\!$2.4$ & \!\!\!$5$ & $6\times 10^{-4}$ & \!\!\!\!$10^{38}$ & \!\!\!\!$1.0 \times 10^{31}$\\
        \!\!\!3 & \!\!\!$2.4$ & \!\!\!$1.25$ & $10^{-2}$ & \!\!\!\!$10^{38}$ & \!\!\!\!$4.4 \times 10^{31}$\\
        \!\!\!4 & \!\!\!$2.4$ & \!\!\!$1.25$ & $0.1$ & \!\!\!\!$10^{38}$ & \!\!\!\!$4.8 \times 10^{31}$\\
        \!\!\!5 & \!\!\!$2.4$ & \!\!\!$1.25$ & $6\times10^{-4}$ & \!\!\!\!$10^{39}$ & \!\!\!\!$3.9 \times 10^{33}$\\
        \!\!\!6 & \!\!\!$2.4$ & \!\!\!$1.25$ & $6\times10^{-4}$ & \!\!\!\!$10^{40}$ & \!\!\!\!$4.2 \times 10^{35}$\\
        \!\!\!7 & \!\!\!$2.0$ & \!\!\!$1.25$ & $6\times 10^{-4}$ & \!\!\!\!$10^{38}$ & \!\!\!\!$4.0 \times 10^{31}$\\
        \!\!\!8 & \!\!\!$1.5$ & \!\!\!$1.25$ & $6\times 10^{-4}$ & \!\!\!\!$10^{38}$ & \!\!\!\!$5.6 \times 10^{31}$\\
        \hline
        \end{tabular}
\label{modelparameters}
        \end{table}
    
\begin{figure*}
    \centering
    \includegraphics[width=0.98\hsize]{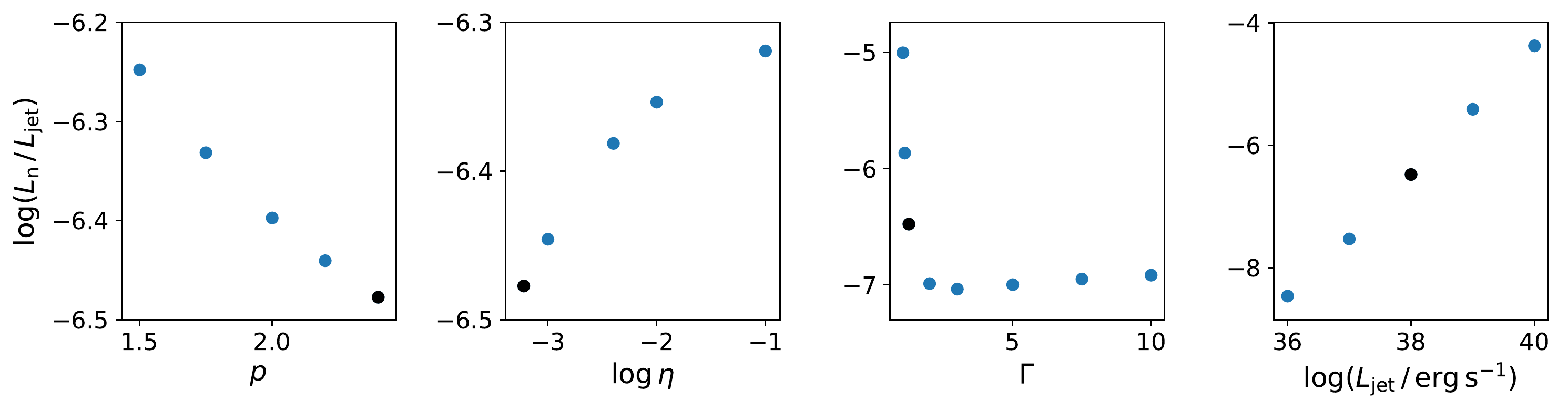}
    \caption{Neutron-to-jet-power ratios for different parameter values. Black dots correspond to the parameters of the \object{Cygnus X-1} model and blue dots to the rest of the models. We perform variations of one parameter in each case, with the remaining parameters fixed to the values of the Cygnus X-1 model. From left to right, the variable parameter is: injection index $p$, acceleration efficiency $\eta$, the bulk Lorentz factor of the jet  $\Gamma$, and its luminosity $L_{\mathrm{jet}}$.}
    \label{Fig:neutronpower_models}
    \end{figure*}

    Figure~\ref{Fig:neutronpower_models} shows the neutron-to-jet-power ratios for a wider range of parameter values. The variation of the spectral index and the efficiency parameter have a mild impact on the total neutron power. For reasonable values of these parameters, the power ratio varies by less than an order of magnitude. On the other hand, we observe a greater effect when varying the Lorentz factor and luminosity of the jet, which produces changes in the neutron power of several orders of magnitude. We note a rapid increase as $\Gamma \to 1$. This is due to the increase in the bulk proton density, which results in a higher $p$-$p$ rate. Therefore, as $\Gamma$ increases, the $p$-$p$ rate decreases, and so does the neutron power. At $\Gamma \approx 3,$ the effect of the Lorentz boost overcomes that of lower neutron production rates, producing a slight increase in the neutron power. Another important result is that the power ratio increases almost linearly with the jet luminosity. In other words, luminous jets are more efficient in transferring energy to the neutron component. This arises because the density of bulk and relativistic protons are both proportional to the jet luminosity, rendering the total neutron power $L_{\mathrm{n}}$ quadratic in $L_\mathrm{jet}$. A general result of this section is that our model predicts that slow, high-luminosity jets are the astrophysical systems in which energy feedback into the ISM by neutron transport may play an important role.

\section{Cosmic-ray production}
\label{sec:cr_production}

\subsection{Neutron decay}      
        
        We consider beta decay of free neutrons, $n \longrightarrow p + e + \bar{\nu}_{e}$ \citep[][\emph{et seqq.}]{Fermi1934}. The decay distance $r$ follows an exponential probability density function $f(r\,\text{;}\gamma_{\mathrm{n}})$ with mean $c\gamma_{\mathrm{n}}\tau_{\mathrm{n}}$, where $\gamma_{n}$ is the neutron Lorentz factor and \mbox{$\tau_\mathrm{n}\approx 881.5\,\mathrm{s}$} 
    is the neutron lifetime \citep{Wietfeldt2018}.
    The power deposited within $r$ in secondary particles is given by

    \begin{eqnarray}
    \label{power_distance}
    P'_\mathrm{d}(r') = 
    \frac{1}{2}\int_0^{r'} \int_0^\infty \int_{-1}^1 P(E_{\mathrm{n}}) \frac{\mathrm{d}\mu}{\mathrm{d}\mu'} \frac{\mathrm{d}E_{\mathrm{n}}}{\mathrm{d}E'_{\mathrm{n}}} f(r'\,\text{;}\gamma'_{n}) \mathrm{d}\mu'\, \mathrm{d}E'_{\mathrm{n}} \mathrm{d}r',
    \end{eqnarray}  
        
\begin{figure}
    \centering
    \includegraphics[width=0.95\hsize]{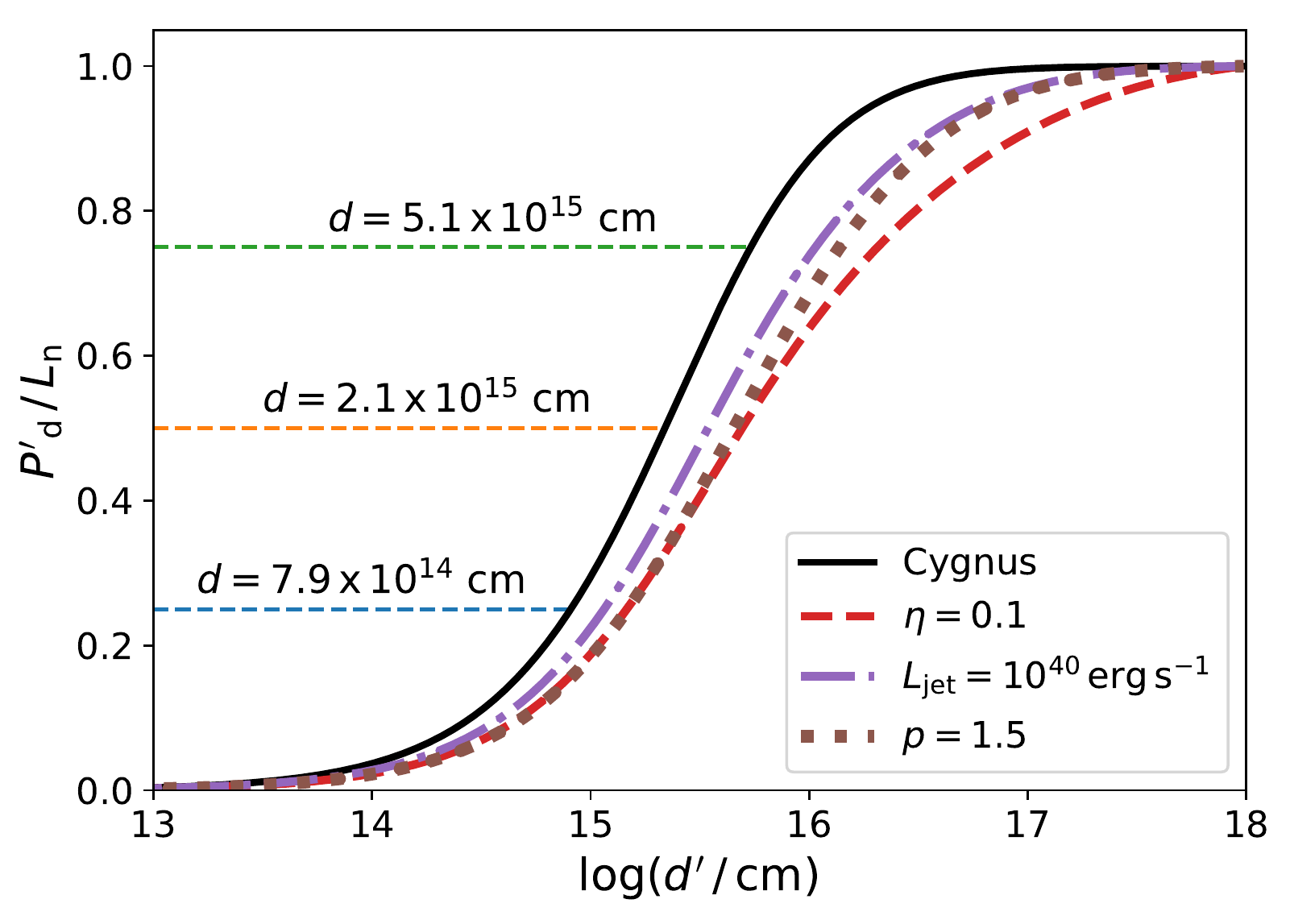}
    \caption{Power injected in secondary particles arising in neutron decay, $P_{\mathrm{d}}$, normalised to $L_{\mathrm{n}}$, up to a sphere of radius $d'$ centred at the MQ. Dashed horizontal lines indicate the first (blue), second (orange), and third quartiles of the total power. Four models presenting different behaviour are shown: \object{Cygnus X-1} (black solid line), and models 4, 6, and 8 (red dashed, lilac dot-dashed, and brown dotted lines, respectively).}
    \label{Fig:InjectionEnergy}
    \end{figure}

        \noindent
        In Fig.~\ref{Fig:InjectionEnergy} we show $P_{\mathrm{d}}$ as a function of distance. 
        We observe that, in our Cygnus X-1 model, most of the power is injected at distances $\gtrsim 10^{15}$~cm and up to $\sim 10^{17}\,\mathrm{cm}$, values which are $\sim 10^{3} - 10^{5}$ times the binary system separation. This distance increases slightly, about half an order of magnitude, for models that produce harder neutron spectra.

\begin{figure}
    \centering
    \includegraphics[width=0.95\hsize]{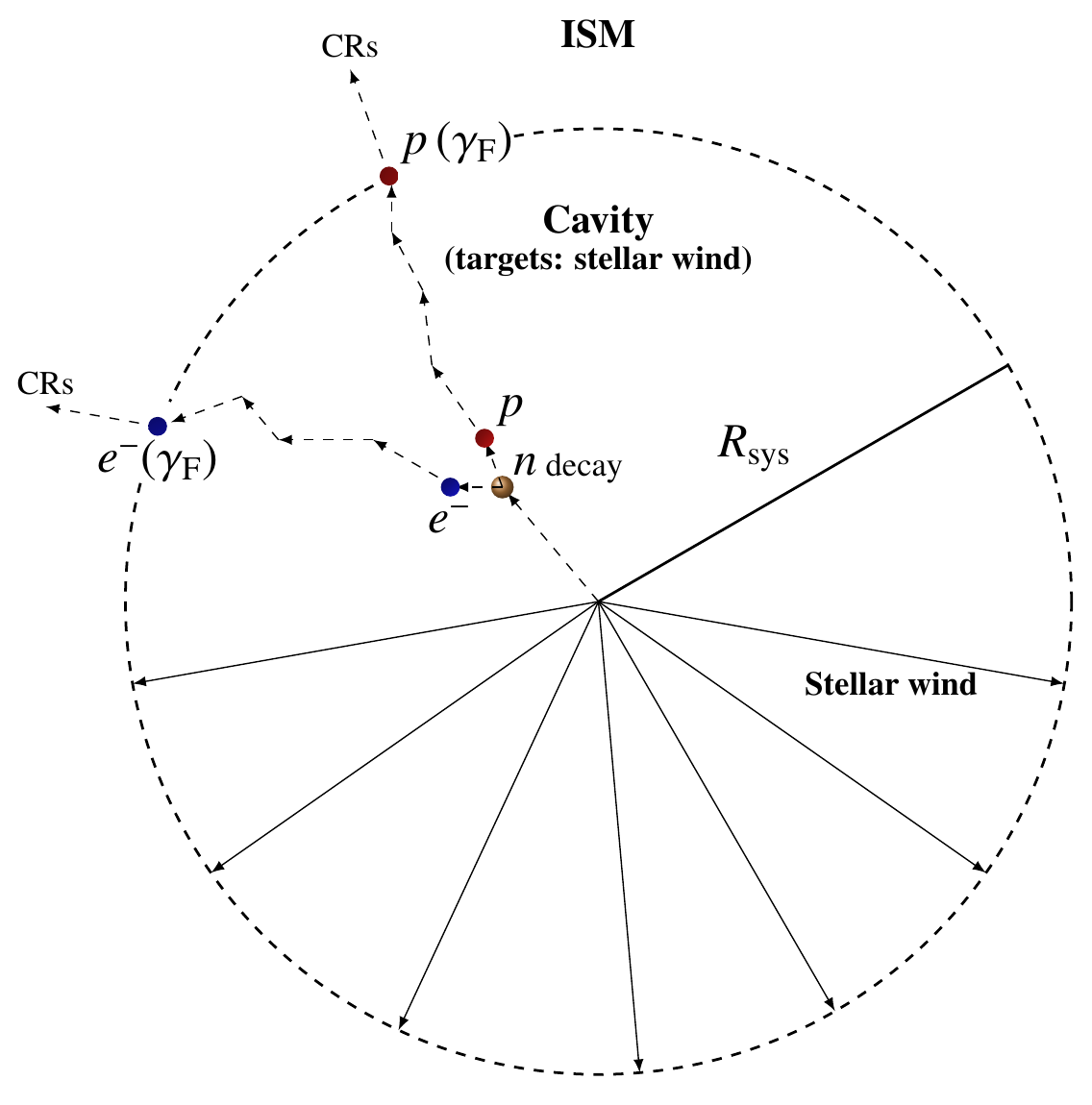}
    \caption{Picture of the transport of particles within the stellar wind cavity (not to scale). The cavity is centred at the MQ. Neutrons propagate radially outwards until they decay into protons and electrons (neutrinos can be neglected for the purpose of this work). Charged particles follow a stochastic motion due to diffusion in the stellar wind plasma, losing energy until they reach the ISM and become CRs.}
    \label{Fig:Cavity}
    \end{figure}

    Secondary particles propagate through the matter, radiation, and magnetic fields surrounding the system.  We consider the stellar wind of the companion as the main matter field in the decay region. This wind will expel the ISM matter and form a cavity of radius $R_{\mathrm{sys}}$,
which is        given by the distance where the wind pressure is equal to that of the ISM (Fig.~\ref{Fig:Cavity}),
        beyond which we assume a typical ISM field. Thereby, we assume that particles that escape from the cavity 
        become CRs. Given the high velocity and mass-loss rate of massive stars, it is expected that $R_\mathrm{sys} \gg R_\star$, and therefore we can take $v_{\mathrm{w}} \approx v_{\infty}$ for the velocity of the stellar wind in that region. Thus, the distance at which both pressures equilibrate is given by
    
        \begin{eqnarray}
    R_{\mathrm{sys}} = \left( \frac{\dot{M} v_{\mathrm{\infty}}}{24 \pi p_{\mathrm{ISM}}} \right)^{1/2},
    \end{eqnarray} 
    
    \noindent where $p_{\mathrm{ISM}}$ is the pressure of the ISM. 
    
    For the stellar wind of \object{HDE 226\,268} (the massive O9.7 star in \object{Cygnus X-1} system) and a typical value of $p_\mathrm{ISM} \approx 10^{-12}~\text{dyn}~\text{cm}^{-2}$, we obtain $R_{\mathrm{sys}}\approx 2.3\times10^{19}~\mathrm{cm}$. Therefore the injected particles propagate inside the cavity formed by the stellar wind before escaping. The same applies for lower mass-loss rates, down to $\dot{M} \approx 10^{-8} ~\mathrm{M}_{\odot} ~\mathrm{yr}^{-1}$, and for the whole range of wind velocities of massive stars \citep[$\sim 100 - 3000\,\textrm{km s}^{-1}$, e.g.][]{2012A&A...541A.145C}. Therefore, in high-mass MQs, neutron products would almost always have to travel some distance to reach the ISM, losing part of their energy.
        
        The energy deposited in neutron-decay products may be carried away from the system by them, radiated through
        their interactions with magnetic, photon, and matter fields, or transferred to the medium by elastic interactions. Adopting typical values for stellar magnetic fields ($B \approx 100$ G for the surface of a high-mass star), synchrotron losses are negligible in comparison to the adiabatic losses that particles suffer when propagating through the wind plasma. Relevant fields for proton--photon or electron inverse Compton interactions come from the binary system (companion star,
        accretion disc, jet, etc.). However,  the collision rate
        is negligible in both cases because the encounter is produced at very small angles ($\lesssim 0.001$), as both 
        colliding particles propagate outwards away from the system. Regarding proton--proton inelastic collisions, the mean free path is $\gtrsim 10\,\mathrm{pc}$ for the wind-matter field.
        On the other hand, for typical values of magnetic field, the electron synchrotron cooling rate implies that their energy is radiated within typical distances $\gtrsim 1~$kpc, depending on the neutron decay distance. Thereby, the emission inside the cavity would be negligible. 
        Thus, radiative losses of these particles are negligible while they propagate towards the ISM. Instead, before emerging as cosmic rays, part of their energy is lost while diffusing through the plasma.
        
   \subsection{Cosmic-ray spectra}      
        
        To compute the losses of secondary particles in the stellar wind until they reach the edge of the cavity, we take the formulae for the same process in the solar wind, given by \citet{Gleeson1978} and \citet{Strauss2011}. In our case, the energy-loss rate can be written as
        
        \begin{eqnarray}
        \label{eq:lorentz-variation}
    \dot{\gamma} = -\frac{2}{3}\gamma \beta^{2} \frac{v_{\mathrm{w}}}{r},
        \end{eqnarray}
    
\noindent where $\gamma$ is the Lorentz factor of the particle and $v_{\mathrm{w}}$ the stellar wind velocity at a given distance $r$ from the binary system. These energy losses are the result of particles propagating diffusively in the cavity through scattering off magnetic waves in the plasma.

    The radial motion equation of relativistic particles is given by $r = \sqrt{D\,t} + r_{0}$, where $r_{0}$ is the injection ---neutron decay--- distance and $D$ is the diffusion coefficient, for which we adopt the Bohm approximation. In terms of the model parameters, $D\approx E\,c\,r^{3}r_{\star}^{-3} / (3eB_{\star})$, where $B_{\star}$ and $R_{\star}$ are the surface magnetic field and radius of the companion star, respectively. We use this relation to integrate Eq.~\ref{eq:lorentz-variation} from the injection position $r_{0}$ to $R_{\mathrm{sys}}$, yielding
    
        \begin{eqnarray}
    \label{eq:escape_energy}
    \gamma_{\mathrm{F}} - \gamma(r_0) + \frac{1}{2}\ln\left( \frac{\gamma_{\mathrm{F}} - 1}{\gamma_{\mathrm{F}} + 1} \right) = - \frac{2}{9} \frac{v_{\mathrm{w}}}{r_0^{2}}\frac{3 e B_{\star} R_{\star}^{3}}{m c^{3}},
    \end{eqnarray}
                
        \noindent where $\gamma_{\mathrm{F}} = \gamma({R_{\mathrm{sys}}})$. The value of $\gamma(r_0)$ depends on the decaying neutron energy, which is distributed among the created proton (99.9\%) and electron (0.1\%). Equation~\ref{eq:escape_energy}
        gives the energy at which particles escape the system and emerge as cosmic rays in the ISM. Using this equation, we compute the cosmic-ray spectra assuming the number of particles is conserved for each population.
        
    \begin{figure}
    \centering
    \includegraphics[width=0.95\hsize]{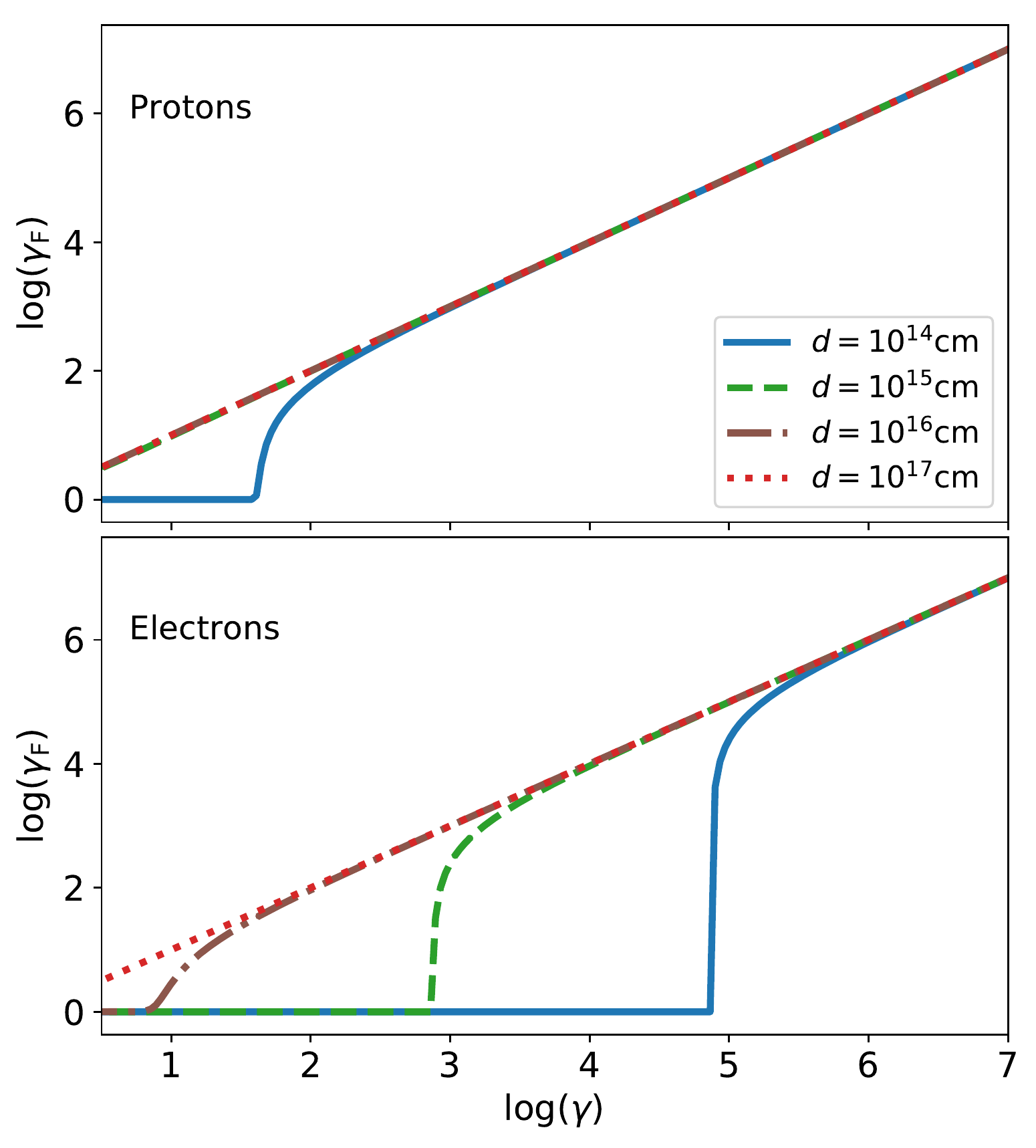}
    \caption{Lorentz factor at escape, $\gamma_{\mathrm{F}}$, vs. Lorentz factor at injection, $\gamma$, for protons (top) and electrons (bottom) injected at different distances: $d = 10^{14}~$cm (blue solid line), $d = 10^{15}~$cm (dashed green line), $d = 10^{16}~$cm (dashdotted brown line), and $d = 10^{17}~$cm (dotted red line). For these results we adopted a typical surface value of $B_{\star} \approx 100$ G for a high-mass star.}
    \label{Fig:finallorentz}
    \end{figure}
    
    \begin{figure}
    \centering
    \includegraphics[width=0.95\hsize]{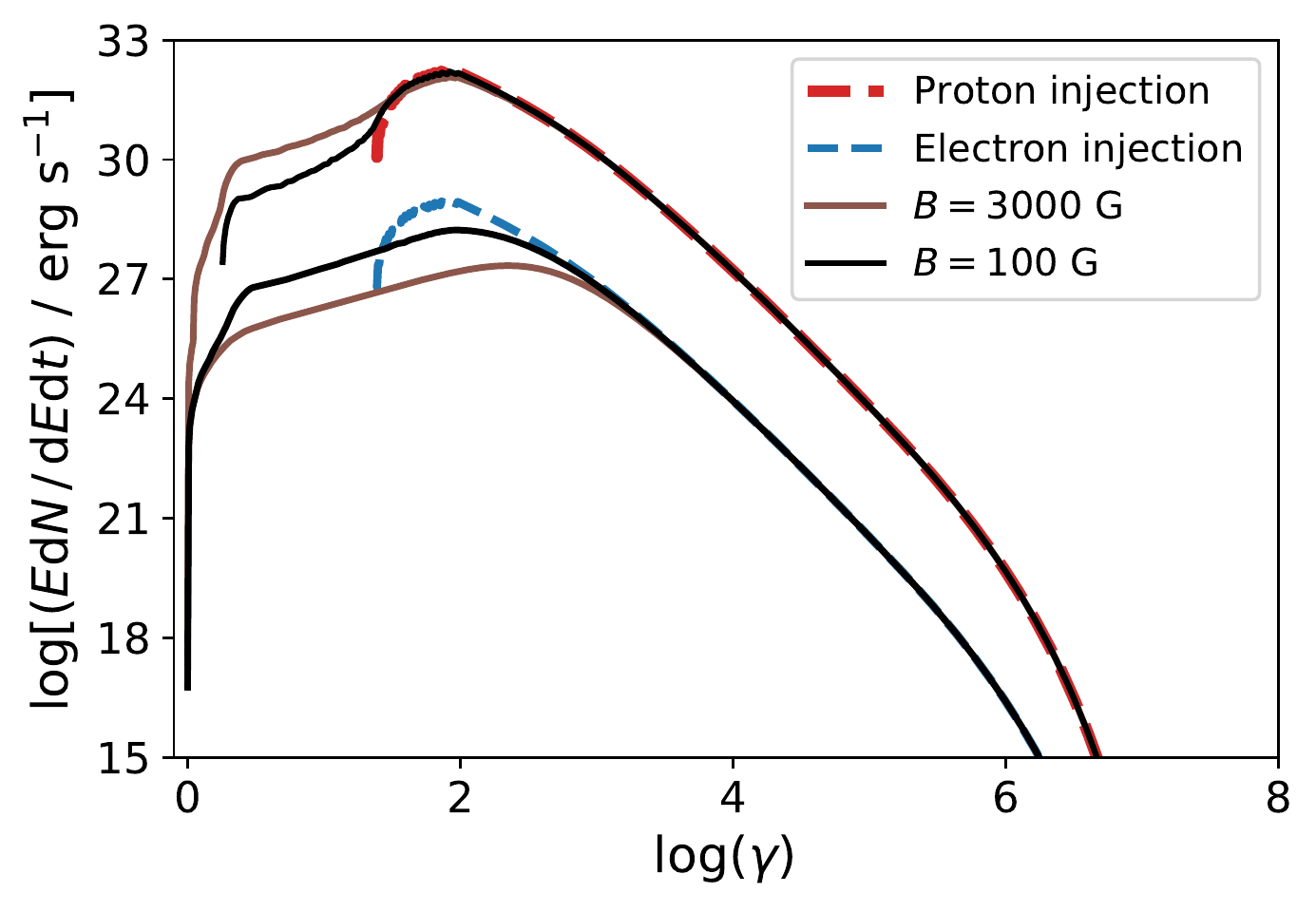}
    \caption{Cosmic-ray spectra produced by neutrons in MQs. Blue and red dashed lines represent injected electron and proton spectra, respectively. Black and brown solid lines represent the cosmic-ray spectra for the case of a companion star with $B \approx 100$ G and $B \approx 3000$ G, respectively. In the case of protons, all spectra are almost identical for energies $E \gtrsim E_{\mathrm{min}}$.}
    \label{Fig:crspectra}
    \end{figure}
    
    Figure~\ref{Fig:finallorentz} shows the final Lorentz factor of secondary particles injected at different distances. Particles coming from neutrons that decay near the source suffer more losses as they diffuse through a longer path. The proton population does not suffer significant losses for decay distances greater than $\sim 10^{14}~$cm. The effect of diffusion
is greater for electrons, even at large decay distances. Figure~\ref{Fig:crspectra} shows the spectra of CRs (protons and electrons) at $r = R_{\mathrm{sys}}$ for our fiducial model. The cosmic-ray proton spectrum is almost the same as the injected one, with a spectral index of $p \sim 3$, but presents a tail at low energies due to diffusion. The electron spectrum flattens as a consequence of electrons suffering higher diffusion losses, and accumulating at lower energies in a tail similar to that of the proton spectrum. Both spectra show a maximum value around, and related to, the minimum Lorentz factor of the proton population. We recall that for our fiducial model, $E_\mathrm{min} = 95.4\,mc^{2}$. For other jet models, this value may vary down to $\sim 2 mc^2$, and consequently, the maximum of the CR spectra will be located at lower energies. 
    
    The total CR spectrum carries essentially all the power deposited in the neutron population. Therefore, for any model explored in previous sections, the total cosmic-ray power is given by $L_{\mathrm{CR}} \approx L_{\mathrm{n}}$ (see Table \ref{Tab:models}). Losses suffered by particles before escaping from the system depend on the wind velocity and injection distance.  Therefore, other parameters will not produce significant variations for these energy losses, and the cosmic-ray spectra will modify analogously to the neutron spectra.
    
\section{Discussion and conclusions}
\label{Conclusions}

We introduced the relativistic neutron component in hadronic jet models through inelastic proton-proton collisions that produce these particles \emph{in situ}. The density of the proton population overcomes that of the neutrons by a factor of $\gtrsim 10^{2}$ and neutron decay is negligible inside the jet. Therefore, the steady-state proton population is almost the same as that obtained without considering neutron production. The same is true for the jet SED, making the identification of the neutron component in MQs by its emission unattainable with present instruments.

Neutrons escape and decay far away from the jet, but remain inside the cavity carved out in the ISM by the MQ companion, as far as high-mass MQs are concerned. These neutrons inject secondary relativistic protons and electrons that propagate diffusively and finally escape from the cavity into the ISM, becoming CRs. These particles do not radiate significantly within the cavity, being undetectable through electromagnetic emission. They may effectively carry a small fraction of up to $10^{-4}$ of the jet power out into the ISM, depending mainly on the jet luminosity and bulk velocity. The distribution of this power among the proton and electron components is governed mainly by neutron decay physics, except by the small amount of power lost by electrons in diffusing out of the system. 

Microquasars have been considered as potential CR sources in previous works \citep[e.g.][]{Heinz&Sunyaev2002}. These authors computed the CR output of cold protons and heavy ions that accelerate and escape through the terminal shock in the jet. The main feature of their CR spectra is a narrow shape around a typical Lorentz factor of at most a few times that of the jet bulk, which is usually $\Gamma \sim 3-10$. Alternatively, our mechanism produces broad spectra peaked at half the minimum energy of relativistic protons in the jet, which may be more than one order of magnitude higher. The shape of the CR spectrum is similar for protons and electrons in our case; it presents an almost flat tail below the peak energy, and a steep decay with an index around $-3$ above. Briefly, our neutron-escape-based scenario can provide more energetic CRs because it drains energy directly from the relativistic proton population at the base of the jet, instead of taking that of cool particles emerging at its end.

It is important to recall that \citet{Heinz&Sunyaev2002} assume that CRs emerging from the terminal shock of the jet are injected directly into the ISM. However, typical jets have lengths of up to $\sim 10^{15}\,\mathrm{cm}$, much smaller than the sizes of the cavities carved out by MQ companions in the ISM. Therefore, it is expected that the spectra of CRs exiting the jet through the terminal shock are modified by the wind. For  particles with small Lorentz factors, such as those obtained by \citet{Heinz&Sunyaev2002}, and stars like the companion of Cygnus X-1, our results suggest that CRs would thermalise within the stellar wind if the terminal shock lies at $\lesssim 10^{14}\,\mathrm{cm}$. Therefore, the CR spectrum may be highly modulated by the wind, depending on the specific properties of the system. This issue merits a more thorough investigation.

Supernova remnants are at present considered the main sources of Galactic CRs. They inject $\sim 10^{51}~$erg into the ISM. However, the efficiency of the acceleration of CRs in SNR shocks remains under discussion. It is accepted that if $\sim 10\%$ of the SNR energy is used in CR acceleration, the supernova population might explain the observed CR power in the Galaxy up to energies lower than that of the knee of the CR spectrum. \citet{Hovey2018} measured an upper limit of $\sim 7\%$ for this efficiency, depending on the ionisation factor of the pre-shock gas, whereas \citet{Shimoda2015} argue that the CR acceleration efficiency may have been overestimated by $10-40\%$. Regarding our results, and assuming that a MQ jet like \object{Cygnus X-1} might be active over a time of $\sim 10^{6}~$yr, it would inject only $\sim 10^{45}~\mathrm{erg}$ into the ISM, far below the contribution of an individual SNR. Even slow jets, such as that of \object{SS 433}, would inject only $~3\times 10^{46}~$erg, still a low CR power. Only ultraluminous X-ray sources with the most powerful stellar jets can compete with SNRs by producing up to $\sim 10^{49}~$erg, about 10\% of the CR power of a SNR. More optimistic scenarios than those mentioned could be obtained combining parameters that favour neutron production independently, but would hardly represent the typical MQ population. It is important to recall that our estimates rely on a conservative value for the branching ratio of the neutron production channel. Other authors adopt values that are  higher  by a factor of up to six \citep[cf.][]{Sikora1989, Atoyan1992a, Atoyan1992b, Vila2014, Romero2020}, which would increase the neutron power by a similar amount.

On the other hand, measurements of the CR spectrum observed at Earth suggest that it is steeper than that predicted by actual models of diffusive acceleration in SNR shocks \citep[e.g.][]{Blasi2013}. In particular, spectral indices softer than $\sim 2$ (the canonical value in the standard theory of diffusive shock acceleration) are required to describe observations. Our results suggest that these values may be easily achieved by a neutron-escape-based mechanism. This is due to a steepening of the neutron spectrum, which is the result of a contribution of several spectra with different values of the maximum energy achieved at each region in the jet (compare with Fig. \ref{fig:neutronpower}).

A recent lepto-hadronic model for Cygnus X-1 was proposed by \citet{Kantzas2021}. The main differences with the model on which ours is based \citep{Pepe2015} are that the acceleration region extends to higher distances along the jet axis, and that synchrotron-self-Compton photons are adopted as targets for proton--photon interactions. This latter interaction is another source of relativistic neutrons that could dominate over the proton--proton channel in some cases. According to our model, the neutrons produced at farther regions would contribute to the lower energies of the population and modify the spectrum accordingly (see Fig. \ref{fig:neutronpower}).

A more precise estimate of the contribution of MQs to Galactic CRs should take into account population considerations, because of the different production rates, lifetimes, and duty cycles of both classes of objects, which depend on the properties of the parent stellar populations. An interesting by-product is that old stellar populations might contribute to Galactic CRs through low-mass MQs. In the case of a low-mass companion star, harder neutron spectra could be expected, because their slow winds produce small cavities, allowing neutrons to inject a significant amount of power directly outside. This may also
be the case for low-metallicity stars, which produce weaker winds. In these cases, the products of neutron decay emerge directly as cosmic rays.
    An exploration of these issues requires the development of stellar evolution models that include the MQ phase, and a more complete census of Galactic MQs to determine their actual population. A rough computation has already been presented by \citet{Fender2005}, who estimate the MQ contribution to the Galactic cosmic-ray population to be in the range of $5-10\,\%$. Similar studies may also shed light onto the population of CRs of star-forming galaxies, and the origin of their $\gamma$-ray emission \citep[see e.g.][and references therein]{Romero2018,Kornecki2020}.

Based on the increase in the XRB production rate and luminosity at low metallicities, \citet{Mirabel2011} proposed that these sources may have contributed to the ionisation and heating of the IGM in the early Universe through their X-ray emission. This argument has been extended to include the contribution of CRs from MQs \citep{Tueros2014,Douna2018}. The latter authors emphasise that the electron spectrum is a key ingredient, finding that MQs with soft electron spectra provide the largest ionisation powers. Our work therefore provides a mechanism to support the claims of \citet{Douna2018}. Moreover, \citet{Sotomayor2019} present theoretical models in which Population III MQs produce CRs in the terminal regions of extremely powerful hadronic jets ($L_\mathrm{jet} \sim 10^{41}\,\mathrm{erg\,s}^{-1}$). Our neutron-escape-based mechanism predicts that those systems would have a very strong CR emission of $\sim 10^{38}-10^{39}\,\mathrm{erg\,s}^{-1}$, and therefore a large ionising and heating power. Taking into account the fact that theoretical models predict that Population III stars have weak winds as a consequence of their low metallicities, relativistic neutrons would decay directly in the ISM, without suffering diffusion losses. Therefore, the contribution of our mechanism to ionising CRs in the early Universe merits exploration. We will present our results on this issue in a forthcoming paper.

\begin{acknowledgements}
The authors acknowledge the anonymous referee for valuable comments that greatly improved the manuscript.
GJE and LJP acknowledge support from project PIP 2014/0265 from Argentine CONICET.
\end{acknowledgements}

%
   \bibliographystyle{aa} 
   \bibliography{bibliography} 
%

\end{document}